\documentclass{PoS}

\newcommand{\three}{\ensuremath{3\times1\times1
    \xspace\mbox{m}^3}\xspace}

\usepackage{silence}
\usepackage{amsthm} 
\usepackage{cleveref}
\usepackage{graphicx}
\usepackage{longtable}
\usepackage{tabularx}
\usepackage{booktabs}
\usepackage{xspace}
\usepackage{lscape}
\usepackage{subfigure} 
\usepackage{setspace}
\usepackage{epsfig}
\usepackage{units}
\usepackage{cleveref}
\usepackage{threeparttable}
\usepackage{multirow}
\usepackage{comment}
\usepackage{float}

\newcommand{\six}{\ensuremath{6\times6\times6 \xspace \mbox{m}^3}\xspace}

\newcommand{\figref}[1]{\figurename~\ref{#1}}

\title{The WA105-3x1x1 m$^3$ dual phase LAr-TPC demonstrator}

\ShortTitle{The WA105-3x1x1 m$^3$ dual phase demonstrator}

\author{\speaker{Sebastien Murphy} on behalf of the WA105 collaboration\\
        ETH Zurich\\
        E-mail: \email{Sebastien.Murphy@cern.ch}}

\usepackage{url}

\abstract{ The dual phase Liquid Argon Time Projection Chamber (LAr
  TPC) is the state-of-art technology for neutrino detection thanks to
  its superb 3D tracking and calorimetry performance. Its main feature
  is the charge amplification in gas argon which provides excellent
  signal-to-noise ratio. Electrons produced in the liquid argon are
  extracted in the gas phase. Here, a readout plane based on Large
  Electron Multiplier detectors provides amplification of the charges
  before its collection onto an anode with strip readout.  The charge
  amplification enables constructing fully homogenous giant LAr-TPCs
  with tuneable gain, excellent charge imaging performance and
  increased sensitivity to low energy events. Following a staged
  approach the WA105 collaboration is constructing a dual phase
  LAr-TPC with an active volume of \three that will soon be tested
  with cosmic rays. Its construction and operation aims to test
  scalable solutions for the crucial aspects of this technology: ultra
  high argon purity in non-evacuable tank, large area dual phase
  charge readout system in several square meter scale, and accessible
  cold front-end electronics.  A milestone was achieved last year in
  the completion of the 24 m$^3$ cryostat that hosts the TPC. This is
  the first cryostat based on membrane technology to be constructed at
  CERN and is therefore also an important step towards the realisation
  of the upcoming protoDUNE detectors.  The \three dual phase LAr-TPC
  will be described in and we will report on the latest construction
  progress.}

\FullConference{38th International Conference on High Energy Physics\\
		3-10 August 2016\\
		Chicago, USA}

\begin{document}

\section{Introduction}
Giant Liquid Argon Time Projection Chambers (LAr TPCs), at the 10 kton
level, are at the design and prototyping stage in the context of the
Deep Underground Neutrino Experiment (DUNE) \cite{DUNE_CDR}.  Liquid
Argon TPCs, provide a complete 3 dimensional image of the neutrino
interaction final state particles overa wide range of energies,
allowing for efficient background rejection and good energy
reconstruction.  The double (or dual) phase liquid Argon
TPC~\cite{Rubbia:2004tz,Badertscher:2010zg,Badertscher:2013wm}
represents a novel concept for liquid argon detectors.  Compared to
the single-phase, the dual-phase design will provide a fully active
volume without dead material, a smaller number of readout channels, a
finer readout pitch, a more robust signal-to-noise ratio with tunable
gain, a lower detection energy threshold, and a better pattern
reconstruction of the events avoiding induction views. These will
allow to best exploit the ``bubble chamber''-like features of the
liquid argon TPC at the 10-kt scale.  The aim of the WA105 experiment
at the CERN Neutrino Platform is to fully demonstrate the dual phase
technology at the scales of the DUNE Far Detector, by constructing and
testing full-scale detector components, assessing their installation
procedures in a 300 ton protoDUNE dual phase demonstrator and to
measure the detector performance in a charged particle test beam
\cite{Agostino:2014qoa}. Following a staged approach, the WA105
collaboration has already constructed a smaller dual phase \three
prototype that is about to be commissioned and will start data taking
with cosmic rays in the coming months. This prototype is the largest
dual phase LAr TPC ever to be built and is the concrete result of many
years of R\&D on smaller detectors.


\section{Overview of the \three detector}

\begin{figure}[h!]
  \centering
  \includegraphics[width=1\textwidth]{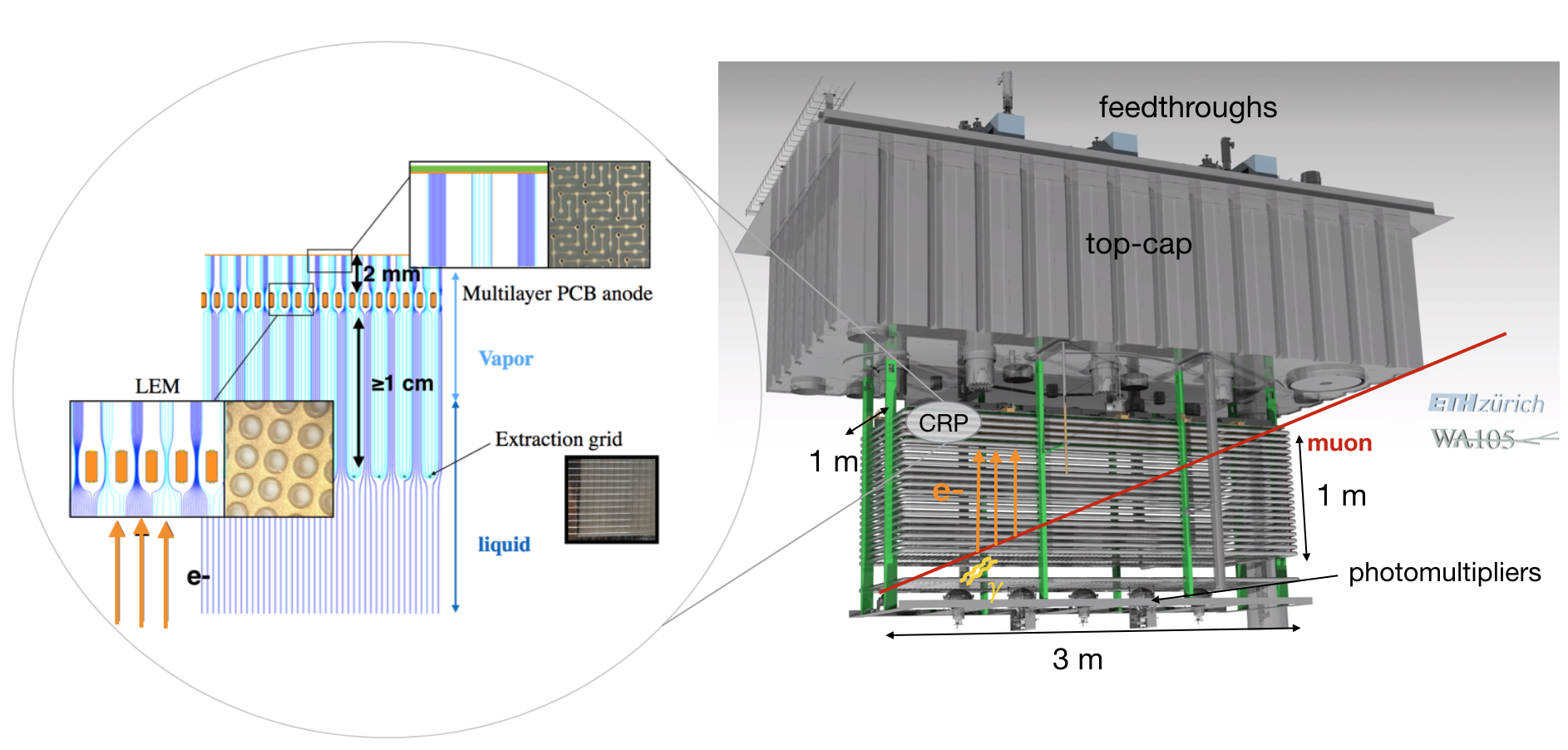}
  \caption{Drawing of the WA105-\three dual phase LAr TPC. The
    zoom provides an illustration of the charge amplification region
    near the surface of the liquid argon}
     \label{fig:311-full} 
   \end{figure}

The WA105-\three dual phase TPC is illustrated in
   \figref{fig:311-full}. It consists of a one meter high field cage
   made by 20 field shapers placed at a constant spacing of 50 mm and
   a metallic grid cathode.

   A uniform drift field is provided by a
   resistor divider chain situated between the cathode and the top
   field shaper. At the top the drifting charges are extracted to the
   gas phase where they are amplified and readout by a 3 $\times$ 1
   m$^2$ charge readout plane (CRP).  Five photomultiplier tubes
   (PMTs) coated with the wavelength shifter, TPB, are fixed under the
   cathode. They are sensitive to the 128 nm scintillation light from
   the argon scintillation and provide the reference time for the
   drift as well as the trigger. The entire detector is hung under a
   1.2 m thick insulating top cap.  The field cage is fixed by eight
   FR4 bars and the CRP is suspended by means of three adjustable
   cables inserted in dedicated suspension feedthroughs. The top cap
   is part of the cryostat structure providing the functionality of
   reducing heat input and minimizing the liquid and gas Argon
   convection. The thermal insulation of both top-cap and cryostat is
   based on GRPF (glass reinforced polyurethane foam) layers,
   interspersed with pressure distributing layers of plywood. Its
   thickness and composition is such to reach a residual heat input of
   5 W/m$^2$ in cold operation. The inside of the cryostat is covered
   with 1.2 mm thick corrugated steel panels that are precisely shaped
   to absorb the thermal contractions.

\subsection{The Charge Readout Plane}
The key concept of the dual phase LAr-TPC, relies on extracting the
ionisation charge to the Argon gas phase where it can be amplified by
Large Electron Multipliers (LEMs)~\cite{Bondar:2008yw}. The LEMs
function by triggering Townsend multiplication in the high electric
field regions inside their holes. The principle is illustrated in the
zoom of \figref{fig:311-full}, the electrons are efficiently extracted
from the liquid with an electric field of around 2 kV/cm and amplified
with a field of about 30 kV/cm applied across both electrodes of the
LEM. 
\begin{figure}[h!]
  \centering
  \includegraphics[width=.9\textwidth]{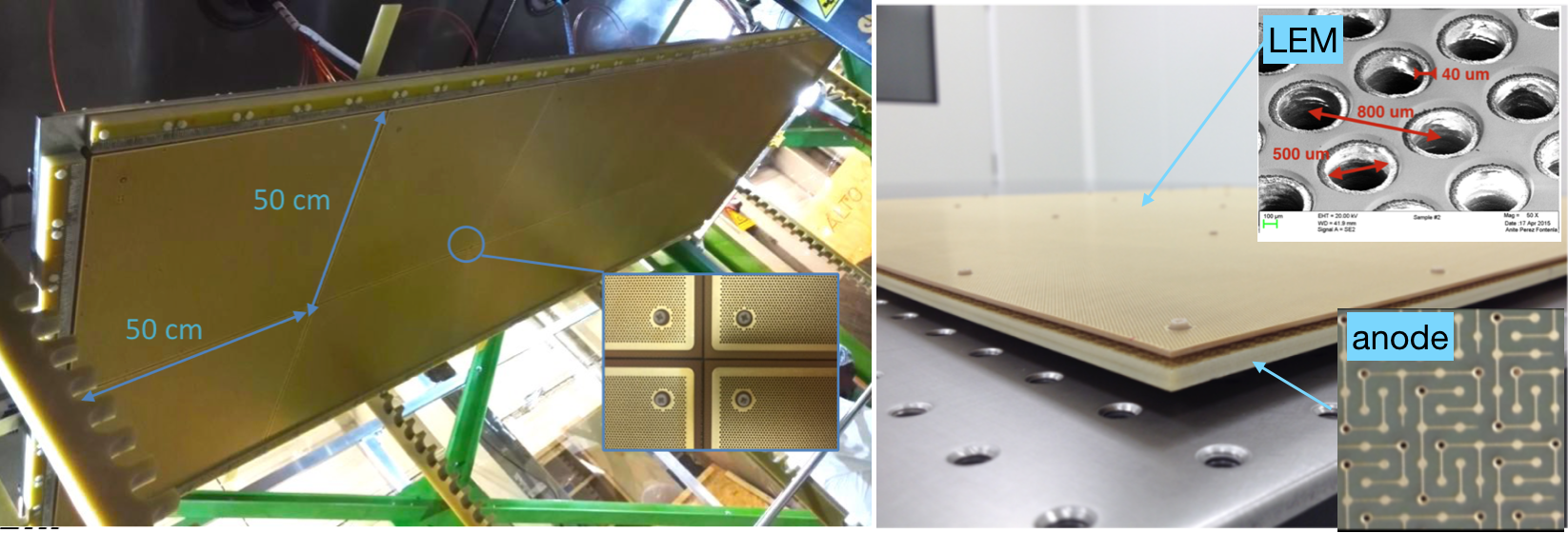}
  \caption{Bottom view of the Charge Readout Plane showing the fully
    active 3$\times$1 m$^2$ amplification area. An individual LEM and anode
    sandwich with zooms on their surfaces is shown on the right. }
  \label{fig:311-CRP} 
\end{figure}
The electric field to extract the charges is provided by an extraction
grid placed 5 mm below the LAr surface. Once amplified in the LEM the
electrons are then collected and recorded on a two-dimensional and
segmented anode. The anode consists of a set of strips (views) that
provide the 2D $x$ and $y$ coordinates of the event with a 3 mm pitch.
Both the anodes and LEMs are rather standard printed circuit boards
(PCBs) that can easily be mass-produced in the industry at affordable
costs. In the design concept of the \three , all those stages
(extraction grid, LEM and anode) are assembled in a single Charge
Readout Plane (CRP). The CRP therefore consists of the 2D anodes, the
LEMs and the extraction grid assembled as a multi-layered ``sandwich''
unit with precisely defined inter-stage distances and inter-alignment.
A picture of the CRP during detector assembly is shown in
\figref{fig:311-CRP}. The extraction grid is made from 100 micron
diameter wires matching the readout pitch of 3 mm and tensed across
the 3$\times$1 m$^2$ area of the CRP . The anodes and LEMs on the
other hand come in individual units of 50$\times$50 cm$^2$ PCBs.
Twelve of each are precisely positioned on the frame to provide a
fully active amplification and readout area.  One of the technological
deliverables of the \three is to demonstrate a uniform gain and
detector response on readouts at the multi-square meter scale.  In
this respect the design as well as the quality control of the LEMs,
anodes and extraction grid are all crucial aspects.  All designs have
matured following many years of prototyping on smaller scale LAr
detectors. For instance the pattern on the anode was optimised to
ensure exact 50:50 charge sharing between both views and the best
resolution on the particle energy loss per unit length
\cite{Cantini:2013yba}.  LEMs of varying hole sizes, hole pitch, rim
sizes or thicknesses have been operated to verify the geometry that
provides the best and most stable gain in dual phase conditions
\cite{Cantini:2014xza}.  We are therefore confident that the
components of the WA105 CRP match our criteria in terms of stability
and charge uniformity. The CRP frame itself was also dipped multiple
times in open liquid Nitrogen baths and precise survey methods such as
photogrammetry were employed to verify the mechanical tolerances in
cryogenic conditions.

\subsection{Feedthroughs and flanges}
\begin{figure}[h!]
  \centering
  \includegraphics[width=.7\textwidth]{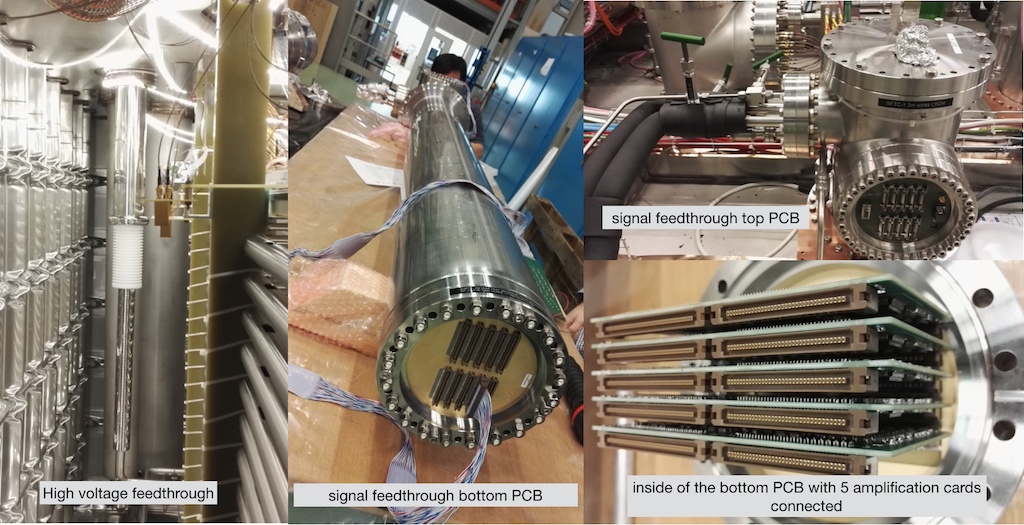}
  \caption{The 300~kV high voltage feedthrough connected to the
    cathode of the detector inside the cryostat (left). A complete
    view of a signal feedthrough (middle) and close up pictures showing
    the top part of the feedthrough as well as the inside of the
    bottom PCB with the preamplifier boards connected.}
     \label{fig:311-FT} 
   \end{figure}

The Feedthroughs are crucial components that serve as interface for
signal, high voltage and slow control sensors between the inside of
the detector and the experimental acquisition.  They are all situated
on the top cap and have to at least penetrate the 1.2 meter insulation
thickness to reach the various parts of the detector (see
\figref{fig:311-full}).  Almost all the feedthroughs installed in the
\three are based on innovative concepts, they will be tested during
the operation and their design will either be copied or scaled up for
future detectors. Since the requirements on vacuum tightness, high
voltage breakdown limits, cryogenic compatibility, are stringent most
of them had to be custom designed in close collaboration with
industry. As example we show in \figref{fig:311-FT} a picture of the
high voltage feedthrough and of one of the signal feedthroughs.

   The role of the high voltage feedthrough is to safely guide the
   high voltage from the power supply cable to the cathode of the
   detector.  It requires a very careful design and precise
   manufacturing to minimise all local electric fields in argon and to
   keep its heat input to a minimum to avoid formation of bubbles. The
   prototype that is installed in the \three was manufactured by the
   company CINEL Strumenti Scientifici s.r.l. It has been successfully
   tested up to $-300$~kV in pure argon in a dedicated setup
   \cite{Cantini:2016tfx}. The ability to withstand a voltage of 300
   kV was guided by the requirements of being able to provide a
   feedthrough for 6 meters drift in the upcoming WA105-\six
   detector. For the \three the feedthrough will be operated at about
   50 kV and will provide feedback on its long term stability.

   An excellent signal-to-noise-ratio is crucial to reach the required
   physics performances, especially for the low energy neutrino
   physics.  In this context an innovative design of signal
   feedthroughs will be tested. They allow to place the amplification
   stage close to the anodes, thereby also profiting from the cold
   environment, while still being able to extract the boards for
   maintenance without accessing the main vessel. Each feedthrough
   reads out 320 channels and consists of a $\sim$ 2 meter long
   stainless steel "chimney" sealed on both ends by circular
   multilayer printed circuit boards (PCBs) with connectors welded on
   both sides. The PCBs are carefully designed to provide ultra-high
   vacuum leak-tightness. The bottom PCB serves as interface between
   the connection to the anode and the five amplifier boards located
   inside the chimney. Each board is guided from the top thanks to
   specially designed FR4 blades. A complete insertion of the blade
   guaranties that the amplifier board is electrically connected to
   the bottom PCB. The top PCB then serves as interface between the
   amplified signals and the digital electronics located on top of the
   cryostat.

\subsection{Slow control sensors}
The slow control system for WA105 is part of a continued progressive
prototyping effort aiming at developing a system dedicated to
multi-ton liquid argon double phase detectors.  It provides precise
monitoring of temperatures, pressures, high voltages and liquid argon
level.  Innovative developments include the construction of very
precise level meters, a motorised suspension system capable of
adjusting the CRP frame with a sub mm precision, special thermometers
to measure temperature gradient in the gas phase, and custom made
cryogenic cameras.
\begin{figure}[h!]
  \centering
  \includegraphics[width=.7\textwidth]{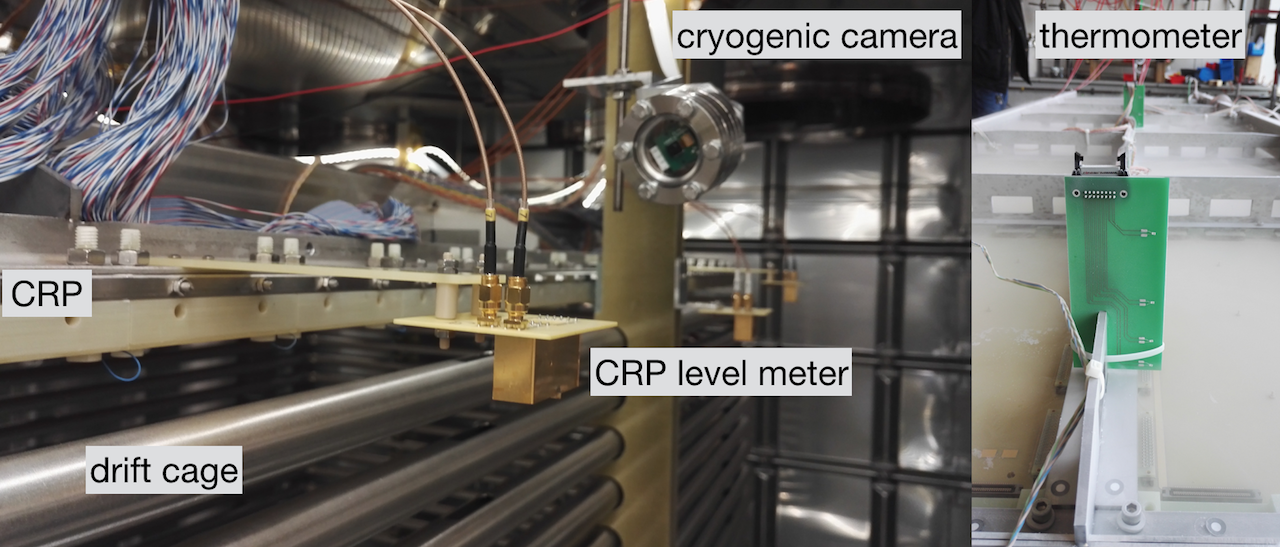}
  \caption{Left: picture of the top of the \three detector inside the
    cryostat. The CRP and part of the drift-cage along with some slow
    control sensors are clearly visible. Right: picture of a
    thermometer during an open cryogenic bath test of the CRP-frame}
     \label{fig:311-SC} 
   \end{figure}
  The cryogenic cameras are essential parts as
   they provide a direct view during operations and are useful for
   instance to monitor the level and stability of the liquid Argon
   level. A picture of some of the sensors installed in the \three is
   shown in \figref{fig:311-SC}. The monitoring system has been
   developed in close collaboration with CERN, a reliable and cost
   effective solution has been found by using National Instruments
   modules mounted in standard experimental racks. This system is
   fully scalable to meet the needs of future larger detectors.


\section{Conclusion}
After more than a decade of R\&D on smal scale prototypes, a first
\three large dual phase liquid argon TPC has been constructed at
CERN. The entire assembly sequence has proven to be straightforward
and rather simple thanks to the attractive design of having a fully
homogenous volume with a minimal number of readout channels. Its
upcoming operation will mark a decisive milestone for the upcoming US
Department of Energy CD-2 in 2019. It is also a first step towards the
realisation of much larger neutrino detectors capable of unprecedented
imaging and compelling physics discoveries such as leptonic CP
violation, the neutrino mass ordering or proton decay.



 \bibliographystyle{utphys}
\bibliography{biblio-ichep}



\end{document}